\documentclass[12pt,a4paper,leqno]{article}
\usepackage{graphicx}
\title{Mathematical models of intergroup conflicts}
\author{S. Panchev$^{*,**}$, Nikolay K. Vitanov$^{***}$}
\date{}
\begin{document}
\maketitle
\begin{abstract}
The human society today is far from perfection and 
conflicts between groups of humans are frequent events. 
One example for such conflicts are armed intergroup conflicts.
The collective behavior of the large number of cooperating 
participants in these conflicts allows us to describe the
conflict on the basis of models containing 
only few variables. In this paper we discuss several cases of  conflicts
without use of weapons of non-conventional kind. 
In the ancient times the  Chinese writer Sun Tsu 
mentioned that the war is an art. We can confirm that the conflict is 
an art but with much mathematics at the background.
\end{abstract}
{\bf Key words}: conflict, attrition, ambush, combat, mathematical models
\section{Introduction}
Today we are excited by the fast advance of the physics and applied 
mathematics in the area of research of complex systems in Nature and society [1-8].
%\cite {grossman, coleman, richards,z1,z2,z3,z4,z5}. 
An important part of the ground for this success was created
more than 60 years ago when L. F. Richardson and L. W. Lanchester, both famous British
scientists, were the first who rose and applied the idea 
for mathematical modeling of arms races and military combats [9-12].
%\cite{r1,r2,r3,r4}.
For many years the research on wars and other military conflicts was concentrated
in the military universities and academies. In the last 20 years and especially 
after the terrorist attack on 11th of September 2001 this research became actual 
for many physicists and applied mathematicians too [13-15].
%\cite{e1,e2,e3}.
\par
In this paper we shall follow the terminology used by Epstein \cite{e4} who applied 
ecological models of Lotka - Volterra kind  for description of  combats.
Let us have two conflicting groups named the "Red group" $R$ 
and the "Blue group" $B$. A general form of model equations of an armed
conflict between these groups is
\begin{equation}\label{model1}
\frac{dB}{dt}=F(B, R;b,r), \hskip.5cm
\frac{dR}{dt}=G(B, R;b;r)
\end{equation}
where $R(t)$ and  $B(t)$ are the  numbers of armed members of the two groups; 
$b$  and $r$ are the "firing efectivenes" (technology level) 
of the groups; and $F$ and $G$ are linear or nonlinear
functions, depending on the character of the conflict. 
Epstein \cite{e4} proposed the following class of models of the conflict
\begin{equation}\label{epst1}
\frac{dR}{dt}= - bB^{c_{1}} R^{c_{2}}, \hskip.5cm
\frac{dB}{dt}= - rR^{c_{3}} B^{c_{4}}
\end{equation}
where $c_{1,2,3,4}$ are  real nonnegative coefficients. If these coefficients are constants
the models are called hard models. If the coefficients depend on the
the number of participants or on the parameters of the environment the
models are called soft ones.
\par
Important characteristics of the model (\ref{epst1}) is the casuality
exchange ratio or state equation (the ratio of eliminated  members of the "red"
 and "blue" groups). For (\ref{epst1}) the ratio is
\begin{equation}\label{cer}
\eta= \frac{dR}{dB} = \frac{b}{r} \frac{B^{\lambda_{b}}}{R^{\lambda_{r}}}
\end{equation}
where
$\lambda_{b}=c_{1}-c_{4}$
and
$\lambda_{r}=c_{3}-c_{2}$.
The integral form of the above casuality exchange ratio is
\begin{equation}\label{icer}
I(B, R) = \frac{b}{\lambda_{b}+1} B^{\lambda_{b}+1} - \frac{r}{\lambda_{r}+1}
R^{\lambda_{r}+1} = I_{0}(B_{0}, R_{0})
\end{equation}
where $B_{0}$ and $R_{0}$ are the numbers of the members of the
two groups at the beginning of the conflict (at $t=0$).
\par
For developing of intuition and decision skills 
it is of interest to know $R(t)$ and $B(t)$ in closed form.
Such analytical solutions are possible only in small number of cases.
Section 2 contains the solution of the system of model equations for selected values
of the parameters $c_{i}$. Several concluding remarks are summarized in section 3.
\section{Analytically solvable models} 
\subsection{The attrition and the square law of Lanchester}
The linear model of Lanchester describes position  conflicts such as the
battles for Somna and  Verdune in 1916. The coefficients in (\ref{epst1})
are $c_{1}=c_{3}=1$, $c_{2}=c_{4}=0$. The equation of state is quadratic
\begin{equation}\label{steq1}
bB^{2}(t) - r R^{2}(t) = b B_{0}^{2} - r R_{0}^{2} = \kappa_{0}
\end{equation}
where $\kappa_{0}$ can be positive, negative, or $0$. For the
case $\kappa_{0}=0$ we obtain $B_{0} = \sqrt{\frac{r}{b}} R_{0}$
and the solutions are
\begin{equation}\label{res1}
R(t) = R_{0} e^{-a t}, \hskip.5cm
B(t) = B_{0} e^{-a t}, \hskip.5cm 
a = \sqrt{br} 
\end{equation}
Evidently $R(\infty) = B(\infty) =0$ which means that after
endless position conflict the two groups are destroyed and no one
of them wins. However the situation changes if $\kappa_{0} \ne 0$.
\par
Let us first assume that $\kappa_{0} >0$. From the state equation (\ref{steq1})
$ B^{2}(t) = \frac{1}{b} (\kappa_{0} + rR^{2}(t))$
and the solutions of the model system of equations are
\begin{equation}\label{solutions1}
R(t) = \frac{c_{0}^{2} \exp(-2at) - \kappa_{0}}{2 \sqrt{r} c_{0} 
\exp(-at)}, \hskip.5cm B(t) = \frac{c_{0}^{2} \exp(-2at) + \kappa_{0}}{
2 \sqrt{b} c_{0} 
\exp(-at)}
\end{equation}
where $c_{0}=\sqrt{r}R_{0} + \sqrt{\kappa_{0}+rR_{0}^{2}} =
\sqrt{r}R_{0} + \sqrt{b} B_{0}$. The obtained solutions satisfy the
initial conditions $R(0) = R_{0}$, $B(0) = B_{0}$. In addition
at $T_{0} = \frac{1}{2a} \ln \left( \frac{c_{0}^{2}}{\kappa_{0}}\right) >0$
we obtain $R(T_{0}) =0$ and $B(T_{0}) = \sqrt{\kappa_{0}/b} \ne 0$. 
In other words the Blue group will win the conflict if it lasts
long enough. The Red group commanders have to change the strategy
if they want to escape the defeat. If this does not happen after the
time $T_{0}$ from the beginning the Red group will be completely 
destroyed - Fig. 1. 
\begin{figure}[t]
\includegraphics[angle=0,width=14cm]{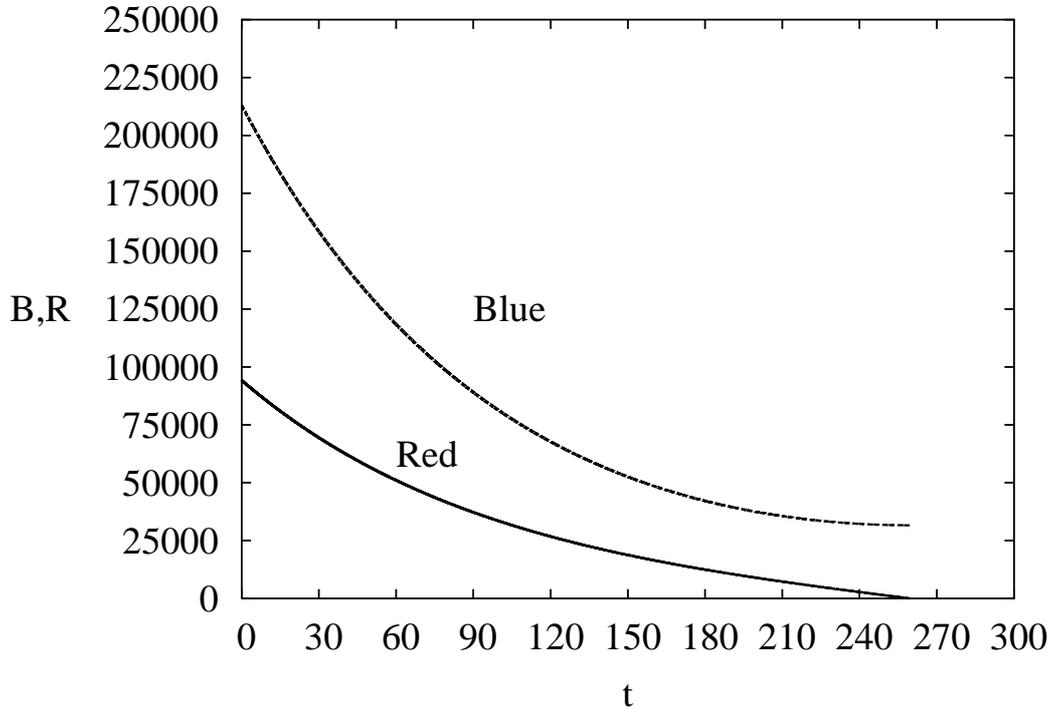}
\caption{Illustration of solution (\ref{solutions1}). The conflict ends
at finite time by destruction of the Red group. Nevertheless the Blue group
suffers heavy losses.}
\end{figure}
We note again that this happens when $\kappa_{0} >0$, i.e., when
\begin{equation}\label{sqlaw}
B_{0} > \sqrt{\frac{r}{b}} R_{0}
\end{equation}
The condition (\ref{sqlaw}) (but with $=$ instead of $>$ ) for military combats
is known as the square law of
Lanchester: to stalemate and adversay army two times as numerous as yours,
your army must be four times as effective. But in this case our army will be also
destroyed. Thus the correct statement of the square law is as follows:
to stalemate and adversay army two times as numerous as yours,
your army must be \underline{{\sl more than}}  four times as effective.
In other words: in position war, in order to stop army that is $n$ time larger
than yours your army has to be more than $n^{2}$ technologically better
(to have more than $n^{2}$ times larger firepower).
\par
We now discuss the case $\kappa_{0} = - \mid \kappa_{0} \mid <0$.
The solutions of the model equations are
\begin{equation}\label{solutions11}
R(t) = \frac{c_{0}^{2} \exp(-2at) + \mid \kappa_{0} \mid}{2 \sqrt{r}
c_{0} \exp(-at)}, \hskip.5cm B(t) = \frac{c_{0}^{2} \exp(-2at) - 
\mid \kappa_{0} \mid}{2 \sqrt{b} c_{0} \exp(-at)}
\end{equation}
Now the Red group wins if
$T_{0} = \frac{1}{2a} \ln \left( \frac{c_{0}^{2}}{\mid \kappa_{0} \mid}\right) >0$
i.e. when $R(T_{0}) =\sqrt{\mid \kappa_{0}/r}$, $B(T_{0})=0$.
\subsection{The concentrated attack model}
In this case the coefficients in the general model are
$c_{1}=c_{2}=c_{3}=c_{4}=1$ and the equation of state is
\begin{equation}\label{steq2}
bB(t) - rR(t) = bB_{0} - rR_{0} = a_{0}
\end{equation}
Epstein assumes that stalemate occurs when 
$ B(t) = R(t)=0$, i.e., when $ rR_{0} = b B_{0}$.
Let us discuss this in more details.
\par
If $a_{0}=0$ (Epstein case) then the solution of the
model system of equations is
\begin{equation}\label{solutions21}
R(t) = \frac{R_{0}}{1+rR_{0}t}, \hskip.5cm 
B(t) = \frac{B_{0}}{1+rR_{0}t}
\end{equation}
thus at $R(\infty)) = B(\infty) =0$, none of the groups wins,
the attack is stopped but the two groups are completely destroyed.
This of course is not of favor for the group leaders. In order
to consider more realistic scenarios we have to set $a_{0} \ne 0$.
In this case the solutions of the model equations are
\begin{equation}\label{solutions22}
R(t) = \frac{a_{0}}{n_{0} \exp(a_{0}t)-r}, \hskip.25cm
B(t) = \frac{a_{0}}{b-m_{0} \exp(-a_{0}t)}
\end{equation}
where $m_{0}=rR_{0}/B_{0}$ and $n_{0}=bB_{0}/R_{0}$. Now the sign
of $a_{0}$ determines the asymptotic behavior of the number of members of the
two groups. If $a_{0}>0$ then $B(\infty) = a_{0}/b = B_{0}-(r/b)R_{0}$
and $R(\infty) =0$. Thus the Blue groups wins and the results of the
attack is that the Red group is defeated -Fig.2  
\begin{figure}[t]
\includegraphics[angle=0,width=14cm]{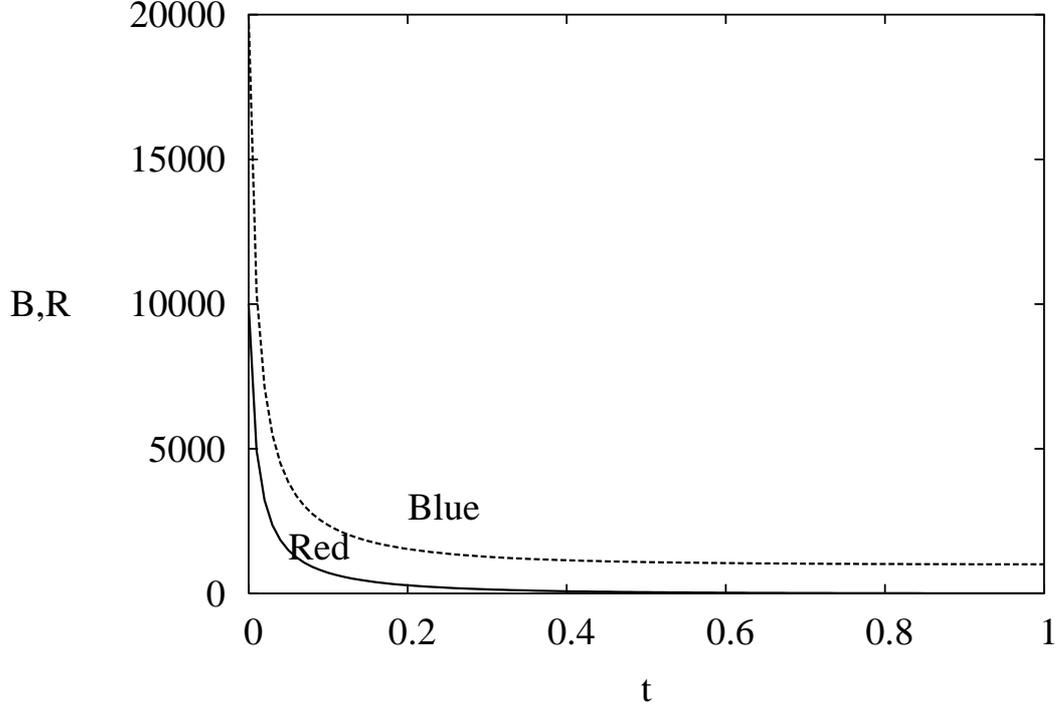}
\caption{Illustration of the solution (\ref{solutions22}). }
\end{figure}
Now let $a_{0} = - 
\mid a_{0} \mid$. Then
\begin{equation}\label{solutions23}
B(t) = \frac{- \mid a_{0} \mid}{b - m_{0} \exp(\mid a_{0} \mid t)}, \hskip.5cm
R(t) = \frac{ \mid a_{0} \mid}{r - n_{0} \exp(- \mid a_{0} \mid t)}
\end{equation}
In this case as winner from the attack scenario is the Red group
as $B(\infty) =0$, $R(\infty) = \mid a_{0} \mid /r = R_{0} - (b/r) B_{0}$.
\par
Now let us discuss the following detail. Let us assume
that at the beginning of the attack the Blue group has more soldiers than
the Red group: $B_{0} >R_{0}$ but the firepower of the Red group is larger:
$r>b$. Then there must be a moment $T_{1}$ where the two groups will have
equal number of armed members: $R(T_{1}) = B(T_{1})$. This moment $T_{1}$ can
be determined from the equation of state (\ref{steq2})
\begin{equation}\label{t2}
(b-r) B(T_{1}) = a_{0}
\end{equation}
What is interesting that when $T_{1}>0$ a solution 
exists only if $a_{0} = -\mid a_{0} \mid <0$  and it is (see Fig.3)
\begin{figure}[t]
\includegraphics[angle=0,width=14cm]{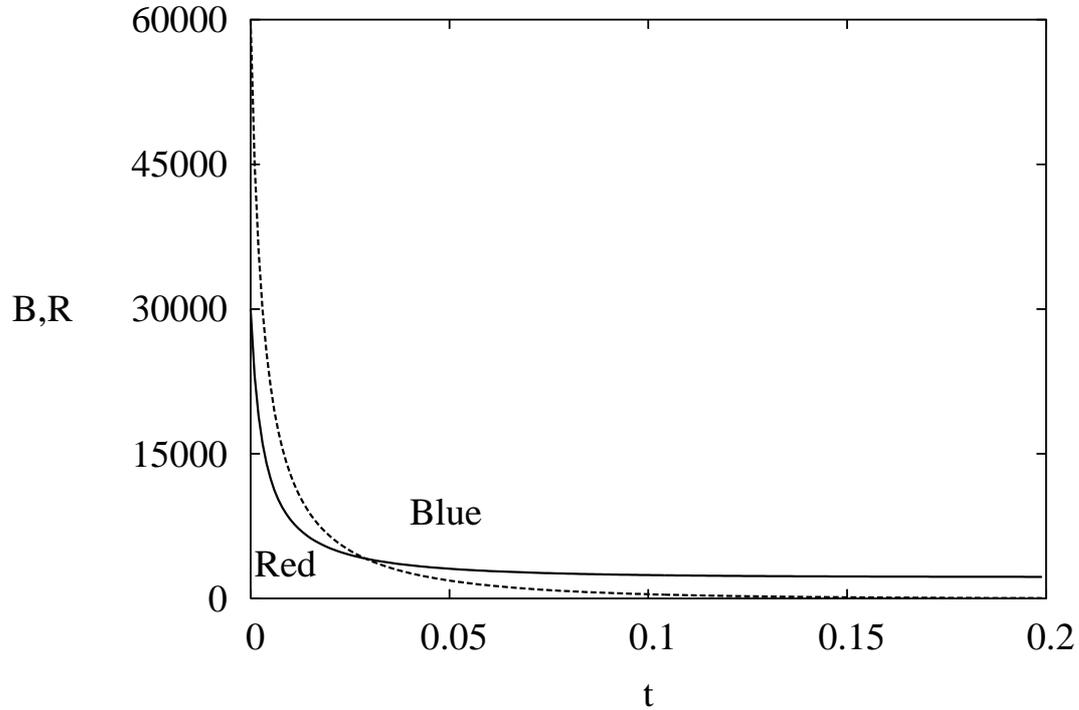}
\caption{Illustration of the solution (\ref{solutions23})}
\end{figure}
\begin{equation}\label{turnpoint1}
T_{1} = \frac{1}{\mid a_{0} \mid} \ln \left( \frac{B_{0}}{R_{0}}\right)
\end{equation}
\subsection{The ambush}
For this case $c_{1}=c_{2}=c_{3}=1$, $c_{4}=0$. The equation of state is
\begin{equation}\label{steq3}
bB^{2}(t) - 2rR(t) = bB_{0}^{2} - 2r R_{0} = s_{0}
\end{equation}
What is interesting here is that the large group does not win
in any case. Let for an example $s_{0}=0$. Then the solutions of the
model equations are
\begin{equation}\label{solution31}
R(t) = \frac{R_{0}}{\left(1+ \frac{1}{2} bB_{0}t \right)^{2}}, \hskip.5cm
B(t)=\frac{B_{0}}{\left(1+ \frac{1}{2} bB_{0}t \right)}
\end{equation}
which means that none wins ($R(\infty)=B(\infty)=0$) and the groups are 
destroyed. This of course is not acceptable for both sides. 
\par
More realistic are situations where $s_{0} \ne 0$. Let first $s_{0}>0$.
The solution of the model equations is
\begin{equation}\label{solution32}
B(t) = \sqrt{\frac{s_{0}}{b}} \frac{1-E_{0} \exp (-\omega_{0} t)}{
1+ E_{0} \exp (-\omega_{0} t)}, \hskip.5cm
R(t) = - \frac{2 s_{0}}{r} \frac{E_{0} \exp (-\omega_{0} t)}{(1+E_{0}
\exp (-\omega_{0} t))^{2}}
\end{equation}
where $ \omega_{0} = \sqrt{ b s_{0}}$ and
$E_{0} = \frac{\sqrt{ s_{0}}- \sqrt{b} B_{0}}{\sqrt{s_{0}} + 
\sqrt{b} B_{0}} <0 $. Thus in the asymptotic case
$B(\infty) = \sqrt{s_{0}/b}$ and $R(\infty) =0$, i.e. the Blue groups
wins.
\par
Quite interesting is the case $s_{0} = - \mid s_{0} \mid <0$. In this case
the solution of the model system is
\begin{equation}\label{solution33}
B(t) = B_{0} \frac{1-\Delta^{-1} \tan (\sigma_{0}t)}{1+ \Delta \tan (\sigma_{0}t)},
\hskip.5cm
R(t) = R_{0} \frac{1+ \tan^{2} (\sigma_{0} t)}{(1+ \Delta \tan(\sigma_{0} t))^{2}}
\end{equation}
where $\sigma_{0}= \frac{1}{2}\sqrt{b \mid s_{0} \mid}$, $\Delta =
B_{0} \sqrt{b/\mid s_{0} \mid}$. In this case at
$ T_{0} = \frac{1}{\sigma_{0}} \arctan (\Delta) $
we obtain $B(T_{0}) = 0, \hskip.5cm R(T_{0}) = \frac{\mid s_{0} \mid}{2r} \ne 0$.
Thus at $t=T_{0}$ this conflict ends. Blue group is completely destroyed
and the Red group wins.
\subsection{Two cubic models}
For these models $c_{1} =c_{3} =2$, $c_{2} = c_{4}=1$ and
$c_{1}=c_{3}=1, c_{2}=c_{4}=2$. Epstein \cite{e4} argues that the
exponents $c_{i}$ in the general model (\ref{epst1}) should be kept in the
interval $[0,1]$. However there is no evidence to sustain such an assertion.
Here we investigate two models for which some of the exponents $c_{i}$ 
equal $2$. 
\par
Let first $c_{1}=c_{3}=2$, $c_{2} =c_{4}=1$. The model equations are
\begin{equation}\label{modeq51}
\frac{dR}{dt}=-bB^{2} R, \hskip.5cm
\frac{dB}{dt}= - rR^{2} B
\end{equation}
The equation of state  is
\begin{equation}\label{steq5}
bB^{2}(t) - rR^{2}(t) = bB_{0}^{2} - rR_{0}^{2} = \kappa_{0} = {\rm const}
\end{equation}
If $\kappa_{0}=0$ then the equation of the state together with one
of the model equations yield
\begin{equation}\label{sol51}
R(t) = \frac{R_{0}}{\sqrt{1 + 2 rR_{0}^{2} t}}, \hskip.5cm
B(t) = \frac{B_{0}}{\sqrt{1+ 2bB_{0}^{2}t}}
\end{equation}
No one of the groups wins at $t \to \infty$. If $\kappa_{0}>0$ we obtain
\begin{equation}\label{sol52}
R(t) = \frac{\sqrt{\kappa_{0}} \exp(- \kappa_{0} t)}{\sqrt{\gamma_{0} -
r \exp(-2 \kappa_{0} t)}}, \hskip.5cm
B(t) = \frac{1}{\sqrt{b}} \sqrt{\kappa_{0} + rR^{2}(t)}
\end{equation}
where $\gamma_{0} =(\kappa_{0} + r R_{0}^{2})/R_{0}^{2}$.
Thus $R(\infty) =0$ and $B(\infty) = \sqrt{\kappa_{0}/b}$.
If $\kappa_{0} <0$ 
\begin{equation}\label{sol53}
R(t) = \frac{\sqrt{\mid \kappa_{0} \mid}}{\sqrt{r+ \gamma_{0} \exp(-2
\mid \kappa_{0} \mid t)}}, \hskip.5cm
B(t) = \frac{1}{\sqrt{b}} \sqrt{r R^{2}(t) - \mid \kappa_{0} \mid}
\end{equation}
Thus $R(\infty) = \sqrt{\mid \kappa_{0} \mid/r}$ and $B(\infty) =0$.
\par
Let now $c_{1} =c_{3} =1$ and $c_{2} =c_{4} =2$. 
The equilibrium condition (\ref{icer}) becomes 
\begin{equation}\label{steq61}
B(t) = B_{0} \left[\frac{R(t)}{R_{0}} \right]^{\epsilon} , \epsilon = \frac{r}{b}
\end{equation}
The final solution is
\begin{equation}\label{sol6y}
R(t) = R_{0} \left(1+ \frac{t}{\tau_{0}} \right)^{-\frac{b}{r+b}}, \hskip.5cm
B(t) = B_{0} \left(1 + \frac{t}{\tau_{0}} \right)^{- \frac{r}{r+b}}
\end{equation}
where $\tau_{0} = 1/((r+b)R_{0} B_{0})$ is a typical time of decay. 
\par
Let us form the ratio
\begin{equation}\label{ratio}
\frac{B(t)}{R(t)} = \frac{B_{0}}{R_{0}} \left(1+ \frac{t}{\tau_{0}} 
\right)^{\frac{b-r}{b+r}}
\end{equation}
where $b>r$ or $b<r$. Assuming $B_{0} >R_{0}$ but $r>b$ and letting $
B(T_{1}) = R(T_{1})$ we obtain 
\begin{equation}\label{tx1}
T_{1} = \tau_{0} \left[ \left( \frac{B_{0}}{R_{0}}\right)^{\kappa} -1\right]
\hskip.5cm
\kappa = \frac{r+b}{r-b}
\end{equation}
Thus $B(t) < R(t) $ at $t>T_{1}$. Nevertheless, both groups,
fight to the end ($B(\infty) = R(\infty) =0$). 
\subsection{A model accounting for epidemic events}
It is known from the history of the conflicts that epidemic events
 had frequently occurred particularly in case of attrition
and ambush conflicts. The simplest way to account for this effect (removing of 
conflict participants because of sickness) is to modify the general model as follows
\begin{eqnarray}\label{dop1}
\frac{dB}{dt}= F(B,R;b,r)-H_{B}, \hskip.25cm H_{B} = k_{B} B, \hskip.25cm k_{B} \ge 0 
\nonumber\\
\frac{dR}{dt}=G(B,R;b,r) - H_{R}, \hskip.25cm H_{R}=k_{R} R, \hskip.25cm k_{R} \ge 0
\end{eqnarray}
where $k_{B}$ and $k_{R}$ are coefficients of morbility (sick rate) removal. Generally
$k_{B} \ne k_{R}$. 
\par
We choose to demonstrate the effect of epidemics on the model (\ref{steq3}) in the form
\begin{equation}\label{dop2}
\frac{dB}{dt} =-rR; \hskip.25cm \frac{dR}{dt}=-bBR - kR, \hskip.25cm k>0
\end{equation}
Hence
\begin{equation}\label{dop3}
bB^{2}(t) - 2rR(t) + 2k B(t) = \tilde{s}_{0} = {\rm const}
\end{equation}
where
\begin{equation}\label{dop4}
\tilde{s}_{0}=bB_{0}^{2} - 2rR_{0} + 2 kB_{0}
\end{equation}
\begin{equation}\label{dop5}
\tilde{s}=bB_{0}^{2} >0 \hskip.25cm {\rm if}\hskip.25cm kB_{0} = rR_{0}
\end{equation}
\begin{equation}\label{dop6}
\tilde{s}=2kB_{0} >0 \hskip.25cm {\rm if}\hskip.25cm bB_{0}^{2} = 2rR_{0}
\end{equation}
In the general case (\ref{dop4}) the quantity $\tilde{s}_{0}$ can be zero, positive or negative. For an example if $\tilde{s}_{0}=0$ the solution of the model system is
\begin{equation}\label{dop7}
B(t) = \frac{2 \epsilon_{0} k \exp(-kt)}{1-\epsilon_{0} b \exp(-kt)}, \hskip.25cm
R(t)= \frac{2}{r} \frac{\epsilon_{0} k^{2} \exp(-kt)}{1-\epsilon_{0} b \exp(-kt)^{2}}
\end{equation}
where
$$
\epsilon_{0}=\frac{B_{0}}{2k+bB_{0}}= \frac{B_{0}^{2}}{2rR_{0}}
$$
These solutions degenerate into (\ref{solution31}) at $k \to 0$.
\section{Summary and conclusion}
A class of mathematical models of armed conflicts (\ref{epst1}) was investigated.
The purpose was to identify particular cases with exact simple solutions in
analytical form $B=B(t)$ and $R=R(t)$, where $B$ and $R$ were the armies' numbers.
It was found that these requirements were met by the linear (Lanchester's) model
known from long ago in the form (\ref{steq1}) as well as by several nonlinear models
described in this paper. These models demonstrate rich behavior in the time.
\par
The predictions of the discussed models fall in two groups
\begin{itemize}
\item
No one of the two groups $B$ and $R$  wins after endless ($t \to \infty$) 
attrition conflict- fighting to the finish (\ref{res1}), (\ref{sol6y}) 
\item
one of the groups wins after limited in time or after an endless conflict (see Table 1)
\begin{table}[t]
\hskip-2.5cm
\begin{tabular}{|c|c|c|c|c|c|} \hline
& & & & & \\
$\frac{dR}{dt}=$ & $-bB$&$-bBR$ & $-bBR$& $-bB^{2}R$& $-bBR^{2}$ \\ 
& & & & & \\
$\frac{dB}{dt}=$& $-rR$&$-rBR$ & $-rR$& $-rR^{2}B$ & $-rRB^{2}$\\ \hline 
 $I_{0}=$& $\kappa_{0}$ ($\dots$)&$a_{0}$ ($\dots$)&$s_{0}$ ($\dots$) &$\kappa_{0}$ ($\dots$), ($\dots$) & $Q_{0}$ ($\dots$)\\ \hline
$I_{0}=0$& $B(\infty)=0$& $B(\infty)=0$&$B(\infty)=0$ &$B(\infty)=0$ &$B(\infty)=0$  \\ 
& $R(\infty)=0$  &$R(\infty)=0$   & $R(\infty)=0$  & $R(\infty)=0$  & $R(\infty)=0$  \\ \hline
$I_{0}>0$& $B(T_{0})=\sqrt{\kappa_{0}/b}$& $B(\infty)=a_{0}/b$&$B(\infty)=\sqrt{s_{0}/b}$ &$B(\infty)=\sqrt{\kappa_{0}/b}$ &$B(\infty)=0$  \\ 
& $R(T_{0})=0$  &$R(\infty)=0$   & $R(\infty)=0$  & $R(\infty)=0$  & $R(\infty)=0$  \\ \hline
$I_{0}<0$& $R(T_{0})=\sqrt{\mid \kappa_{0} \mid/r}$& $R(\infty)=\mid a_{0} \mid/r$&$R(T_{0})=\mid s_{0} \mid /(2r)$ &$R(\infty)=\sqrt{\mid \kappa_{0}\mid/r}$ &$B(\infty)=0$  \\ 
& $B(T_{0})=0$  &$B(\infty)=0$   & $B(T_{0})=0$  & $B(\infty)=0$  & $B(\infty)=0$  \\ \hline
$T_{0}$& $T_{0}=\frac{1}{2a} \ln \left( \frac{c_{0}^{2}}{\kappa_{0}}\right)>0$& $T_{0}=\infty$ &$T_{0}=\frac{1}{\sigma_{0}}\arctan (\Delta)$&$ T_{0}=\infty$ & $T_{0}=\infty$  \\ 
& $c_{0}=\sqrt{r} R_{0}+ \sqrt{b} B_{0}$ & &$\Delta = B_{0} \sqrt{b/\mid s_{0} \mid}$  & &\\ \hline
\end{tabular}
\caption{Summary of model predictions.}
\end{table}
\end{itemize}
Both conflicting groups are defined by their initial numbers $B_{0}$, $R_{0}$ and respective
firing efectivenesses per shot. The character of the conflict is modeled by the form of
the functions $F$ and $G$ (\ref{model1}) which can be linear or nonlinear. All models can
be extended to account for occuring of epidemic events in the fighting groups.
The model (\ref{dop2}) is an example.

\begin{center}
{\sl $^{*}$ Central Laboratory for \\
Solar Terrestial Influences of \\
Bulgarian Academy of Sciences \\
Akad. G. Bonchev Str., Bl. 3 \\
1113, Sofia, Bulgaria\\
e-mail:spanchev@phys.uni-sofia.bg} 
\end{center}
\begin{center}
{\sl $^{**}$ Faculty of Physics \\
"St. Kliment Ohridsky" University of Sofia \\
J. Bourchier 5 Blvd.\\
1164 Sofia, Bulgaria}
\end{center}
\begin{center}
{\sl
$^{***}$ Institute of Mechanics \\ 
Bulgarian Academy of Sciences\\
Akad. G. Bonchev Str., Bl. 4 \\ 
1113 Sofia, Bulgaria}
\end{center}
\end{document}